\documentclass[12pt,preprint,useAMS,a4]{aastex}

\usepackage{natbib}

\usepackage{amssymb}
\usepackage{rotating}

\RequirePackage{lineno}

\usepackage{txfonts}

\usepackage[usenames]{color}

\usepackage{gensymb}

\newcommand{\be} {\begin{equation}}

\newcommand{\fermi}{{\em Fermi}}

\newcommand{\bc}{\begin{center}}
\newcommand{\ec}{\end{center}}
\def\ltsima{$\; \buildrel < \over \sim \;$}
\def\lsim{\lower.5ex\hbox{\ltsima}}
\def\loe{\lower.5ex\hbox{\ltsima}}
\def\gtsima{$\; \buildrel > \over \sim \;$}
\def\gsim{\lower.5ex\hbox{\gtsima}}
\def\goe{\lower.5ex\hbox{\gtsima}}

\def\ltsima{$\; \buildrel < \over \sim \;$}
\def\lsim{\lower.5ex\hbox{\ltsima}}
\def\loe{\lower.5ex\hbox{\ltsima}}
\def\gtsima{$\; \buildrel > \over \sim \;$}
\def\gsim{\lower.5ex\hbox{\gtsima}}
\def\goe{\lower.5ex\hbox{\gtsima}}

\def\ergscm2 {erg\,s$^{-1}$cm$^{-2}$}

\def\cm2 {cm$^{-2}$}

    {\pagebreak\global\pdfpageattr\expandafter{\the\pdfpageattr/Rotate 90}}%
    {\pagebreak\global\pdfpageattr\expandafter{\the\pdfpageattr/Rotate 0}}%

\shortauthors{Torres}
\shorttitle{Order parameters for the high-energy spectra of pulsars}

\begin{document}

\title{Order parameters for the high-energy spectra of pulsars\footnote{Reduced version of an Article published in Nature Astronomy on February 12, 2018. The full  version additionally contains further discussion, tables, and Figures, a section on 
Methods, and Supplementary Materials and it is accessible 
via http://dx.doi.org/10.1038/s41550-018-0384-5 or via SharedIT at http://rdcu.be/GMQm
}}

\author{Diego F. Torres\altaffilmark{1,2,3}}

\altaffiltext{1}{Institute of Space Sciences (ICE, CSIC), Campus UAB, Carrer de Magrans s/n, 08193 Barcelona, Spain}
\altaffiltext{2}{Instituci\'o Catalana de Recerca i Estudis Avan\c{c}ats (ICREA), E-08010 Barcelona, Spain}
\altaffiltext{3}{Institut d'Estudis Espacials de Catalunya (IEEC), 08034 Barcelona, Spain}

\maketitle

{\bf 
From the hundreds of gamma-ray pulsars known \citep{atwood09}, only a handful show non-thermal X-ray pulsations  \citep{kuiper15}.
Instead, nine objects pulse
in non-thermal X-rays but 
lack counterparts at higher energies. 
Here, we present a physical model for the non-thermal emission of pulsars above 1 keV. 
With just four physical parameters, we fit the spectrum of the gamma/X-ray pulsars 
 along seven orders of magnitude.
We find
that
all detections can be encompassed in a continuous variation of the model parameters, 
and pose that their values could likely relate to 
the closure mechanism operating in the accelerating region.
The model explains the appearance of sub-exponential cutoffs at high energies as a natural consequence of synchro-curvature dominated losses, 
unveiling that curvature-only emission may play a relatively minor role --if any-- in the spectrum of most pulsars. The model also explains
the flattening of the X-ray spectra at soft energies as a result of propagating particles being subject to synchrotron losses all along their trajectories.
Using this model, we show how observations in gamma-rays can predict the detectability of the pulsar in X-rays, and viceversa.

}

Curvature and synchrotron emission from particles accelerated in magnetospheric gaps or reconnection (see e.g., 
\citealt{cheng86a,cheng86b,romani96,zhang97,hirotani99a,muslimov03,dyks03,kalapotharakos12,Phil2017})
has been thought to be 
behind non-thermal pulsations.
However, 
a simplified, unifying interpretation of pulsar spectra is still lacking.
Ab initio modelling with sophisticated electrodynamics are not yet focusing on precisely reproducing spectra,
and studies of specific pulsars, with different degrees of complexity, usually produce
unsatisfactory fits and are based on a significant number of case-by-case assumptions
(see e.g., \citealt{kalapotharakos17,hirotani15,takata17}). 
As a result, we normally fit observational spectra just with phenomenological functions:
a power law with a cutoff in gamma-rays, or a log parabola from X-rays up.
This status deters population analysis 
as well as the extraction of physical characteristics.

The basis of the physical model we use here has been presented in a series of papers dealing 
with the limitations and assumptions of earlier outer gap descriptions  \citep{paper1,paper2},
and 
with a systematic analysis of the gamma-ray results obtained  by {\it Fermi}-LAT 
 \citep{paper3,paper4}. 
The model we propose is conceptually simple: it goes from particle dynamics to radiation. 
It is based on the assumption, see e.g., \citet{chiang94,zhang97,kalapotharakos12}, that somewhere near the light cylinder of a pulsar 
having period $P$ and period derivative $\dot P$,
there is a gap with a significant component of electric field parallel to the magnetic lines. This field accelerates particles.
Once a  particle enters into this gap (for instance, by being created there in a pair-production process) the model follows its time-evolution solving the equations of motion, balancing acceleration and losses. This balance takes into account the variation in time/position of the physical
properties along the dynamic propagation of particles, and the influence of 
synchro-curvature radiation (as opposed to the use of synchrotron or curvature radiation only).
With regards to our formerly-quoted series of works, and besides other technical improvements, here we extend the model in two directions. 
On the one hand, we now calculate emission and losses of particles at much lower energies; equivalently, at smaller distances from their injection point. Whereas this does not imply a new physical ingredient per se, it requires the model to cope with a much larger dynamic range, spanning at least 8 orders
of magnitude in position and energy (i.e., at least 3 and 5 orders of magnitude beyond our previous studies, respectively).
This allows us to compute the X-ray emission.
On the other hand, we here pay special attention to the determination of the gradient of the magnetic field in the gap, associating a model parameter to it.
We find that this is essential to model the X-ray data properly and discover how the magnetic field along the gap influences the high-energy predictions too.

\begin{center}
\begin{figure*}
\centering
\includegraphics[scale=.8]{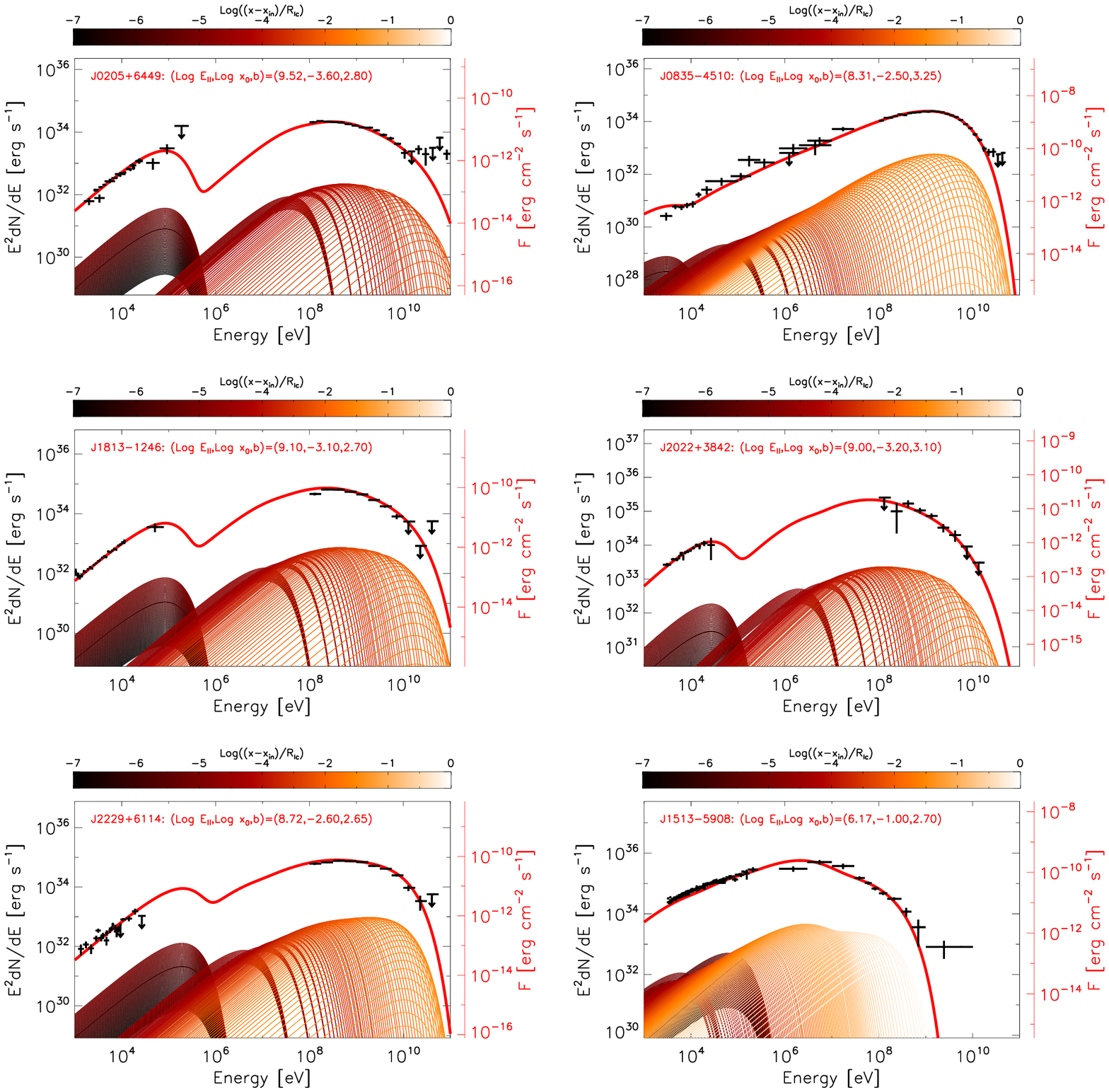}
\caption{
{\bf Results of the model fits for pulsars detected both at X-rays and at gamma-rays. }
Each curve represents the contribution of a given portion of the particles' trajectory along the gap
extent, whose position is color-coded. In order to make the figure legible, not all spatial bins actually computed are shown.
The sum of all contributions constitutes the final fit, depicted by the red curve. $E_{||}$ is given in units of V/m, whereas $x_0$ is given in units of $R_{lc}$.
References for the data used are given in the Methods section.
}
\label{form}
\end{figure*}
\end{center}

We fit observational spectra varying at most four free parameters: 
accelerating electric field,
contrast (a measure of how uniform is the distribution of particles emitting towards us),
magnetic gradient (a measure of how fast the magnetic field declines along the particle trajectory), and
normalization (determining the absolute level, not the shape, of the spectrum). 
Examples of our results are shown in Fig. \ref{form}. 
We find that X-rays are produced by synchrotron-dominated emission 
at the initial loss of the angular momentum of particles after their injection, 
along the first 10$^{-5} R_{\rm lc}$ or less of the extent of their travel ($R_{\rm lc}$ is the pulsar light cylinder).
Instead, regions of the trajectory between $\sim$10$^{-5}$ and $\sim$10$^{-1} R_{\rm lc}$, where 
synchrotron-domination gives pass to a full synchro-curvature regime where
neither synchrotron nor curvature 
significantly dominates 
each other, are responsible for the MeV to GeV emission.

Whereas a change in the magnetic field gradient
does not significantly affect the spectrum in the 0.1--1 GeV band,
it does so at lower and higher energies, and provides a telltale signature in the region of 100 keV -- 10 MeV.
If 
the magnetic field decreases faster, 
the sooner synchrotron-dominated emission ends.  
A larger gradient increases the yield beyond 1 GeV, since
smaller synchrotron losses along the particle movement  allow particles to have a larger Lorentz factor in regions where curvature emission dominates.
In addition, the larger the accelerating field is, 
the larger the peak
energy of the emitted spectrum is.
Thus, for some values of 
magnetic gradient and accelerating field, synchro-curvature emission can  produce sub-exponential fall-offs of the spectra beyond 1 GeV.
This
explains many such instances found in the gamma-ray pulsar population; curvature-only emission cannot produce such sub-exponential cutoffs.

\begin{sidewaysfigure}
\centering
\includegraphics[scale=.8]{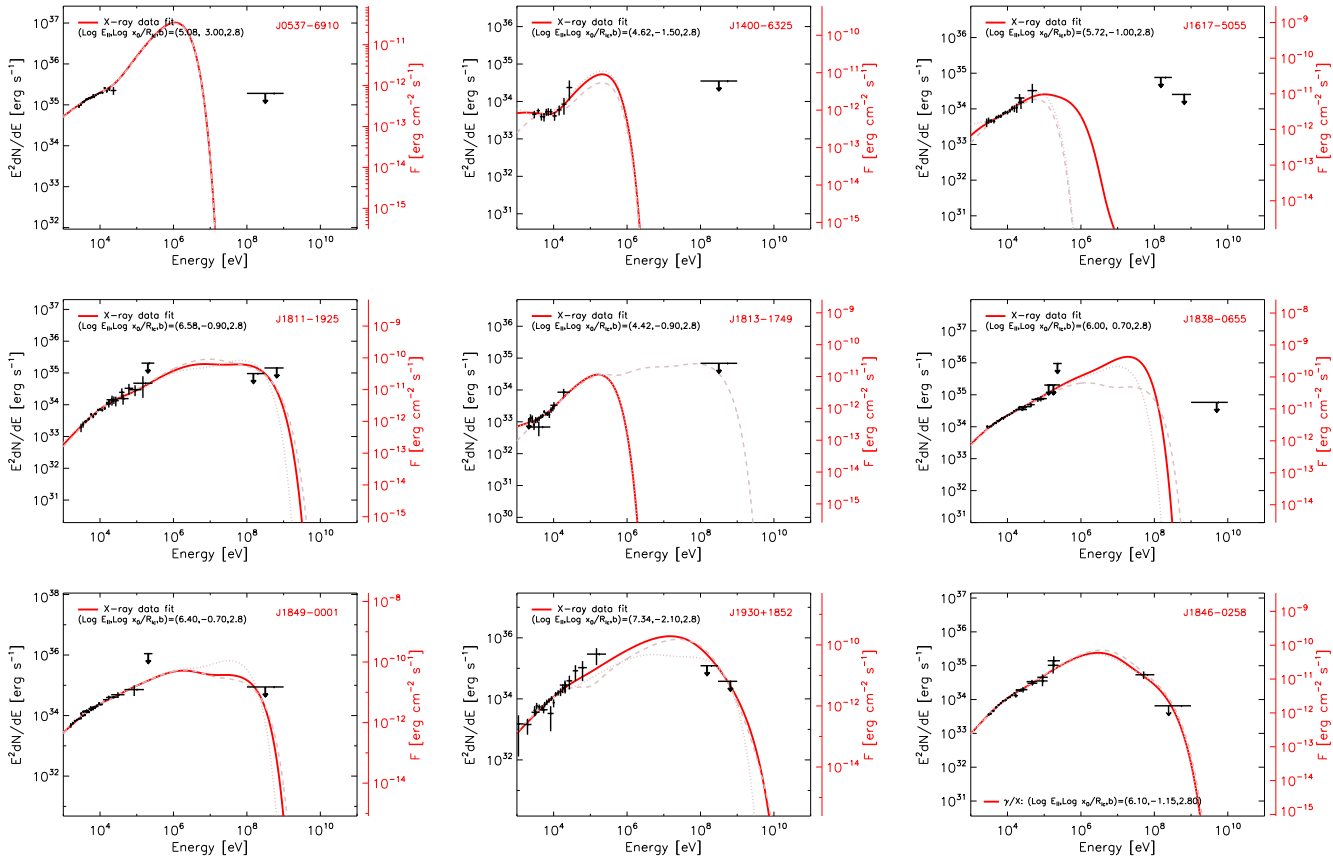}
\caption[angle=90]{{\bf 
Model fits for the pulsars detected only in X-rays. }
The model fits are represented with solid lines, with the best-fit parameters noted in each panel. 
The dashed/dotted curves range on the uncertainties. For details see Methods.
%
In some cases, these  curves 
are very close / superimposed to the best-fits in the scale plotted.
 The last panel shows the results for J1846-0258, for which there is a 4$\sigma$-detection at high-energies.
 Except for this case, the model fits only against the pair ($E_{||},x_0/R_{lc}$) and a global
scaling (with fixed $b=2.8$, and an equal range of the fit parameters in all cases). 
References for the data used are given in the Methods section.
}
\label{x-fits2}
\end{sidewaysfigure}

We now consider objects that are part of the soft gamma-ray pulsar catalog \citep{kuiper15} but have not been detected 
in gamma-rays (Fig. \ref{x-fits2}).
Differently to gamma/X-ray pulsars, 
lower values of accelerating field are generally selected in the fits.
Despite the larger uncertainties produced by the lack of gamma-ray data, we expect PSR J0537-6910, J1400-6325, and J1617-5055 to be GeV quiet pulsars. 
Instead, further studies of 
J1930+1852,  J1811-1925, and perhaps of J1849-0001 and J1838-0655
can lead to a gamma-ray detection. 
These pulsars would be similar to PSR J1513-5908, which has already been detected in gamma-rays.

PSR J1846-0256 (see last panel of Fig. \ref{x-fits2}) was just recently 
claimed  to be a MeV-peaking gamma-ray pulsar at the $4\sigma$-level  \citep{kuiper16a,kuiper17}. The model provides a fit
to the multi-wavelength spectrum much in the same way than it does for PSR J1513-5908.
In fact, considering just the X-ray data, the model predicts that PSR J1846-0256 should be a MeV pulsar, 
likely detectable by Fermi. This provides theoretical support to
the claimed detection 
as well as sustains the predictive power of the model.

\begin{center}
\begin{figure*}
\centering
\includegraphics[scale=.8]{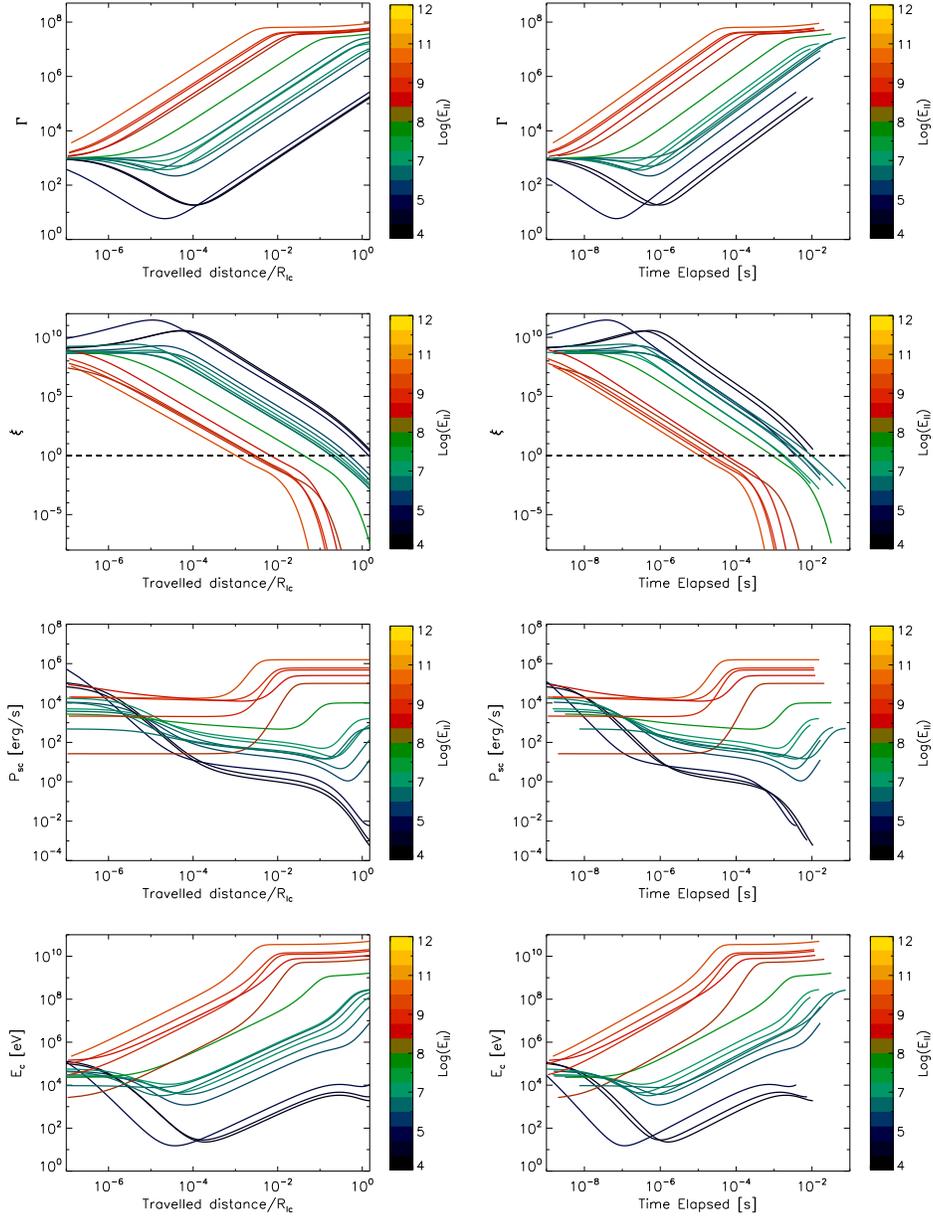}
\caption{{\bf Properties of the trajectories of particles.}
Trajectories of particles as a function of the travelled distance 
(left)  and time (right) for different values of the
accelerating electric field (color-coded logarithmically).
From top to bottom
we show: Lorentz factor $\Gamma$, synchro-curvature parameter $\xi$ (see
Eq. 11 of Methods), synchro-curvature power
($\int ({dP_{sc}}/{dE}) dE$,  see Eq. 3 of Methods),
and characteristic energy $E_c$ (see Eq. 7 of Methods).
The region
of transition between synchrotron ($\xi \gg 1$) and curvature ($\xi \ll1$) domination locates a few orders of magnitude around $\xi=1$.
In the case of X-ray pulsars, we fix the magnetic gradient to $b=2.8$ as in Fig. \ref{x-fits2}.  }
\label{traj}
\end{figure*}
\end{center}

Fig. \ref{traj} shows the dynamical variables along the trajectories related to the fitting solutions. 
Gamma/X-ray pulsars have the larger accelerating fields:
The region of transition between synchrotron and curvature domination determines their spectral shape (see Fig. \ref{traj} together with \ref{form}).
The Lorentz factor of particles  reaches saturation after $\sim 10^{-2} R_{\rm lc}$: 
curvature losses balance acceleration, and the energy remains 
constant afterwards.
This is never achieved in the trajectories of several of the X-ray-only pulsars. 
In fact, for some, 
the (synchrotron-dominated) losses are so high at the 
beginning
that they 
overcome the accelerating force and the Lorentz factor decreases.
For these pulsars, the bins at the farthest gap positions also contribute to the X-ray emission, not to gamma-rays, since the whole particle trajectory is synchrotron-dominated. 
This explains why the observed X-ray spectrum of e.g., J1400-6325,
flattens at low X-ray energies.
In all plots of Fig. \ref{traj},
and running along a decreasing accelerating field, there is a clear distinction between gamma/X-ray pulsars, MeV-peaking pulsars, and X-ray-only pulsars.

We see no particular distinction when
comparing distances and timing properties  of pulsars detected only in X-rays with those detected in both X- and gamma-rays
(see Methods for the computed timing magnitudes). However,
the fact that the fits, all done using the same range of parameter variations, selected a much lower accelerating field and contrast
for some of the X-ray-only pulsars is intriguing: If we consider just the X-ray data for the gamma/X-ray detected
pulsars, the fitted accelerating field and contrast are always larger than those derived for the X-ray-only pulsars, promoting the plausibility of a physical difference. 
A smaller contrast indicates both
that the lightcone of emitted radiation is wider (what is consistent with synchrotron radiation dominating the emission) and 
that a larger extent of the gap is generating particles emitting in the direction of the observer. Both effects imply a larger normalization, what we also find.
It is natural to think that if there are more particles, the screening increases, and the electric field in the gap is reduced. 

The basic mechanism sustaining a gap relies on the acceleration of particles. Since the emitting particles are relativistic, moving out of the gap (either outwards or towards the stellar surface), a feeding mechanism for these particles is needed to sustain the gap existence (this is called {\it closure}). 
The production of pairs is thought to play this role.
For instance, 
gap produced gamma-rays interacting with stellar X-ray photons (either from the neutron star cooling or from bombardment of the flowing backcurrent) can sustain the gaps.
It would not be surprising if the temperature of the blackbody permeating the gap is higher the smaller the peak of the spectral energy distribution is.
With a higher temperature, the luminosity of the soft photons would increase and the pair production opacity would be significant already
for gamma-rays of lower energy. 
However, as soon as the accelerating field is too low, the gap closure mechanism based on pair production of star-generated photons with gamma-rays is in question. 
Fig. \ref{traj} shows that in some cases, the product $E_c E_t$, where $E_c$ is the characteristic energy of the synchro-cruvature radiation and $E_t$ is the energy associated with ($<1$ keV) thermal emission from the star, is sub-threshold for the production of pairs. 
If the cooling or bombardment photon temperature is $<1$ keV,
the soft photons that sustain the pair production cannot come from the star surface.
One could entertain that a special geometry configuration explains why X-ray-only pulsars do not shine in gamma-rays while still have plenty of (invisible) high-energy particles in their magnetospheres \citep{wang13,Wang14}. 
However, if all can be reduced to geometrical effects, 
why all pulsars that were detected in non-thermal
X-rays have the highest ($L_{sd}>5\times10^{36}$ erg s$^{-1}$) rotational power \citep{Wang14,kuiper15}?
An alternative gap closure 
could be that pairs are produced in $\gamma\gamma$ processes in which both photons are synchrotron-generated.
This is energetically allowed by the pair production process, but implies that the energy of the pairs at injection 
must be lower.
We tested the influence of this effect in the spectra for the pulsars with low accelerating field and verified that with 
initial Lorentz factors between 2 and 20 there is no qualitative change to the predicted spectrum: particles are quickly accelerated producing X-rays, spectral peaks in the MeV range, and 
no high-energy gamma-ray emission. Fit parameters comparable to the ones shown in Fig.  \ref{x-fits2} are found within uncertainties.
Under this hypothesis, the more energetic pulsars will be more favourable to sustain gaps via synchrotron photons
when there are no gamma-rays available.
Their
synchrotron luminosity will also scale up, and may be
providing  {in situ}, numerous X-ray photons 
for the production of pairs, leading to a uniform distribution. 
Proving that this is indeed the gap closure mechanism in these cases would however require 3D pair production simulations.

The difference between the spectral fits of the gamma/X-ray pulsars that arise when considering the gamma-ray data only or the whole dataset consisting of both gamma and X-ray data is small:
It is
within a factor of $\sim$2 at 10 keV and lower energies,
for all cases but Vela, for which it is at most a factor of $\sim$10 (see Methods). 
Thus, a model fit to the gamma-ray data 
is a relatively reliable estimator of the non-thermal X-ray detectability, if not the spectrum itself.

We can thus take all gamma-ray pulsars known (i.e., those that lack an X-ray detection) 
and assuming a fixed value of magnetic gradient (we take $b=2.85$, the average of the best-fit $b$-values for detected gamma/X-ray pulsars),
compute a spectrum from 1 keV up: Such spectra can be useful for selecting 
the best X-ray targets for observation. 

\begin{center}
\begin{figure*}
\centering
\includegraphics[scale=.8]{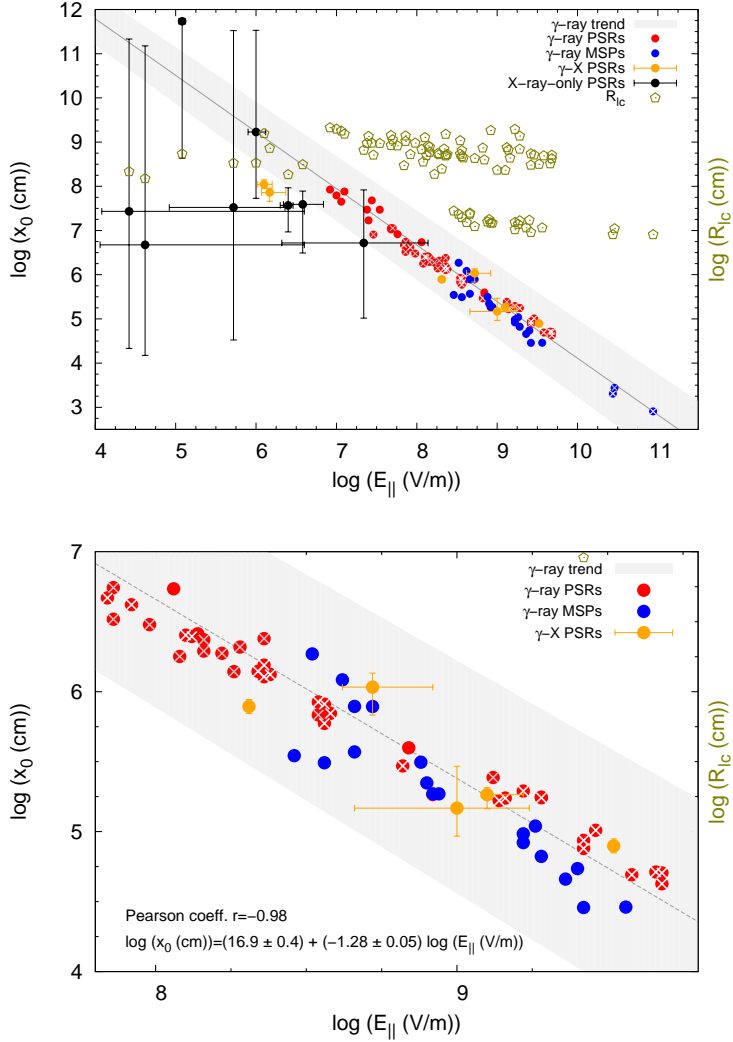}
\caption{ {\bf Correlation of model parameters fitting the high-energy spectra of pulsars and expectations for X-ray detections.} Red and blue dots stand for 
gamma-ray-only, normal and millisecond pulsars, respectively. They show a strong correlation depicted with a dashed line
(the shadowed region represents the 2$\sigma$ uncertainty in this correlation, see Methods). 
This line is extrapolated to lower accelerating fields and its parameters are shown in the second panel, zooming in the central part of the first.
White (grey) crosses within a red/blue colored
point denote a predicted flux of at least 10$^{-13}$ erg cm$^{-2}$ s$^{-1}$ (10$^{-14}$ erg cm$^{-2}$ s$^{-1}$) at 10 keV under the assumed average gradient $b=2.85$.
Further analysis of these expectations is provided in Methods.
The parameters of the X/gamma-ray and X-ray-only pulsars are shown with orange and black points, respectively, together with their 1$\sigma$ uncertainty. 
The light cylinder $R_{lc}$ of all pulsars is also noted (green pentagons). Uncertainties in the model parameters are larger (smaller) when only X-ray (both
X- and GeV gamma-ray) data are available.}
\label{sum}
\end{figure*}
\end{center}

Fig. \ref{sum} shows these predictions   in the context of the 
correlation between the accelerating electric field and the contrast (equal to $R_{lc}/x_0$).
Fig. \ref{sum}  is consistent with our earlier gamma-ray-only study\cite{paper4}, but improves it (apart from using the average $b$-value obtained here from the gamma/X-ray pulsar fits for all gamma-ray only pulsars) by simultaneously considering the X-ray data when available and by recomputing 
the fits for the pulsars detected only in gamma-rays also from 1 keV up.
There is a clear correlation between the model parameters of GeV detected pulsars, 
both millisecond and young, which is confirmed by the cases in which both X-ray and GeV gamma-ray data exist.
A posteriori,
it is possible to understand the correlation in Fig.  \ref{sum} as a confirmation that it is indeed 
synchro-curvature emission what dominates the GeV emission in most gamma-ray pulsars.
Further discussion is provided in the full version of the article.

Future observations with current and forthcoming high-energy missions will sample the physical parameters that yield to very different emission spectra. 
The proposed European eASTROGAM or NASA's AMEGO/ComPair missions \citep{ASTRO,Compair} would be able to detect pulsars 
peaking in the MeV range, like PSR J1513-5908 or J1846-0258, in a region critical for determining their whole non-thermal spectra. 
Detecting pulsed MeV radiation might become common,  unveiling a larger population of pulsars that would have gone unnoticed by \fermi-LAT. 
At the same time, any such  mission would improve the INTEGRAL sensitivity at $>$100 keV by orders of magnitude, so that 
already-detected pulsars will also have their spectrum better defined than what is available today, measuring their peaks and 
removing the uncertainties in the fit parameters. 
The planned eXTP mission \citep{XTP}, and Athena \citep{Athena}, 
could significantly constrain the spectrum of the many gamma-ray pulsars that could be similar to Vela, with a large magnetic gradient in their magnetospheres. 
Those pulsars would have gone undetected yet
(Vela is, because of its proximity), since these would be low-luminous pulsars in X-rays.

\newpage

\subsection*{Acknowledgments}

I acknowledge support from the the grants  
AYA2015-71042-P, SGR 2014--1073 as well as the CERCA Programme of the Generalitat de Catalunya, and the Chinese Academy of Sciences, Grant 11473027.
I am grateful to L. Kuiper, H. An, D. Smith, and J. Li for providing observational data, and
to J. Li, A. Papitto, J. Pons, and D. Vigan\`o for comments.

\bibliography{og}

\end{document}